\documentclass[12pt]{article}
\usepackage{latexsym,amsmath,amssymb,amsbsy,graphicx}
\usepackage[cp1251]{inputenc}
\usepackage[T2A]{fontenc}
\usepackage[space]{cite} 
\usepackage[centerlast,small]{caption2}%

\makeatletter\def\@biblabel#1{\hfill#1.}\makeatother

\allowdisplaybreaks \multlinegap0pt 
\begin {document}
{\flushleft } \noindent\begin{minipage}{\textwidth}
{\small Astrophysics {\bf 63,} 440--446 (2020)\,\,\,\,\,\,\,\,\,\,\,\,\,\,\,\,\,\,\,\,\,\,\,\,\,\,\,\,\,\,\,\,\,\,\,\,\,\,\,\,\,\,\,\,\,\,\,\,\,\,\,\,\,\,\,\,\,\,\,\,\,\,\,\,\,\,\,\,\,\,\,\,\,\,\,\,\,\,\,\,\,\,\,\,\,\,\,\, doi:\,10.1007/s10511-020-09648-x \\
{\scriptsize{\copyright\,2020. Springer Science+Business Media, LLC.
Translated from Astrofizika (August 2020).}}}
\begin{center}
\bigskip
{\normalsize{\bfseries ON THE ORIGIN OF OPTICAL RADIATION DURING THE
IMPULSIVE PHASE
OF FLARES ON dMe STARS.\\
II. CONTINUUM AND LINE RADIATION}}\\[7pt]
\end{center}
{\par \hspace*{1.2cm}\bfseries E. S. Morchenko} {\par
\hspace*{1.2cm}\footnotesize{Physics Faculty, M. V. Lomonosov Moscow
State University, Moscow, Russia;\\
\hspace*{1.1cm} e-mail: \texttt{morchenko@physics.msu.ru}}}
{\par \hspace*{1.2cm}\scriptsize Original article submitted December 16, 2019; accepted for publication June 24, 2020}\\

\setlength{\leftskip}{1.2cm} { \footnotesize{\it It is argued that
not only the blue (at the brightness maximum) but also the red (in
the slow decay phase) components of the optical continuum of
powerful flares on dMe stars are formed near the photosphere. The
possible appearance of HeI lines in the relaxation zones for the
plasma  to a state of thermal equilibrium (as a result of a rise in
electron temperature owing to elastic collisions of electrons with
atoms and ions) is noted (for sufficiently high speeds of a
non-stationary chromospheric shock wave propagating toward the
photosphere of the Sun and stars). A scheme is proposed for the
positioning of the ``layers'' of plasma responsible for the
generation of  radiation in the white light continuum during the
impulsive phase of powerful stellar flares. }}

{\footnotesize{ Keywords: {\it red dwarf stars: flares: the flare
models: optical radiation}\vspace{1pt}\par}}
\bigskip
\bigskip
\end{minipage}\\
{\small{\bfseries{1. Introduction.}} This article completes the
discussion begun by the author in Ref. 1.\footnote{Here (as well as
in \cite{Mor19}), the ``blue'' continuum is considered as {\it part}
of the flare continuum with a sufficiently large optical depth
beyond the Balmer jump. This opacity is a necessary condition
\cite{Grin77} for explaining the {\it blue} color of powerful
stellar flares at the brightness maximum. From a physical
standpoint, the ``blue'' continuum is the same as the
``[quasi-]black-body component'' in terminology Kowalski {\it et
al.} \cite{Kow10} (including the contribution in the $U$ band), as
it follows from Fig. 2 in \cite{Mor15} or identical Fig. 1 in
\cite{Mor16}.}\\
{\bf 2. Radiation from near-photospheric layers.} Belova and Bychkov
\cite{Bychkov19} assume that ``black body radiation that appears at
times during flares'' …
 ``comes from a photosphere heated by a flux of suprathermal particles;'' ``a hot spot with a temperature of $10^4\div2\cdot10^4\text{ K}$ may be formed in it''
  \cite{Bychkov19}.

In these assertions the authors \cite{Bychkov19} do not take the
following into account.  By definition, the visibility of flares
indicates the existence of unperturbed layers of a star's
atmosphere. In the method of Ref. 3 for estimating the area $s$ of
the source of a (quasi-black-body) ``blue'' continuum of powerful
flares on dMe stars (the brightness maximum) in the approximation of
an absolute black body (used, for example, in Ref. 4) these layers
are identified with the photosphere of a red dwarf in a quiescent
state, and the photosphere radiation field is assume to be
Planckian.\footnote{We note that the values of $s$ estimated in Ref.
4 are reduced by a factor of 2. This is because of an imprecision in
Eq. (3); the luminosity of a compact formation radiating according
to the Planck law is equal to \cite{Grin77}
$2\pi{}s{}B_\nu(T_{bb})$, where $T_{bb}$ is the temperature of an
absolute black body corresponding to the spectrum of a flare in the
wavelength $\lambda$ range from 4000 to 4800 \AA.}
   In addition, recent observations
   \cite{Pavl} of the flare activity of Proxima Cen (dM5.5e)
   indicate that the molecular absorption spectra of the atmosphere
   of the red dwarf (the reversing layer) ``are essentially independent of the flare events.'' In turn,
Honda {\it et al.} \cite{Honda} did not detect changes in the
molecular bands during the time of a flare on EV Lac (dM4.5e).

The quasi-black-body shape of the blue continuum of powerful flares
on dMe stars at the brightness maximum is indicated by the existence
of continuum optical radiation on the long-wavelength (red) side
from the $\mathrm{H_\beta}$ line, this radiation does not correspond
to a Planck function with $T_{bb}\sim10^4\text{ K}$ (see Fig. 8a in
the article by Kowalski  {\it et al.} \cite{Kow13}).

Such radiation originates in deeper, less heated near-photospheric
layers. Here an important role in generating the quasi-black-body (a
small Balmer jump) ``red'' component which predominates in the
optical continuum of flares during the slow decay phase (Fig. 31 in
Ref. 7; the wavelengths excluding ``the
conundruum=conundrum+continuum'') is played \cite{Grin77,Grin15} by
the negative hydrogen ion (for $T_{bb}$ approximately $<8000\text{
K}$ the donors of free electrons are metals; see Ref. 9, §6.5). In
terms of this interpretation, the ``relationship'' \cite{Kow13} of
the blue and red components of the continuous radiation of flares,
as well as  the ``relatively more important'' \cite{Kow13}
contribution of the red component ``to the energy budget during the
later stage of the slow decay phase'' \cite{Kow13} seems entirely
natural.

Thus, not only the blue but also the red components of the optical
continuum of powerful flares on dMe stars are formed {\it near} the
photosphere \cite{Grin77}, while the assertions of the authors of
Ref. 2 (and, as a whole, the tendency to ``invoke'' ``layers'' of
the photospheres of dMe stars for interpretation of observations of
flares) conflict with observational data.

The rapid change in the optical continuum of near-photospheric
layers during the impulsive phase of flares makes it difficult to
study gas dynamic processes taking place in {\it higher lying}
layers of the chromosphere since the radiation field of the
near-photospheric gas does influence \cite{Mor15,Mor17CSc} the
ionization state of these layers. Thus, for $T_{bb}=1.1\text{ eV}$
($\approx1.3\cdot10^4\text{ K}$), the electron density
$n_e=10^{14}\text{ cm}^{-3}$, the electron temperature $T_e=1\text{
eV}$, the rate of photoionization by the diluted (dilution factor
W=0.5) Planckian radiation field from the second level of the
hydrogen atom is $\approx4.6$ times greater (\cite{Mor17CSc}, Table
3.2) than the corresponding electron impact ionization rate (results
obtained for the  brightness maximum of a powerful flare; the {\it
approximation} for an absolute black body used here is correct
because the quasi-black-body ``blue'' continuum at the peak of the
flare makes a significant contribution to the radiation including in
the $U$ band; see section 3.3 in the paper by Kowalski {\it et al.}
\cite{Kow10}). At the same time, the rise in $T_e$ behind the front
of the {\it stationary} radiative shock wave owing to elastic
collisions of electrons with atoms and ions for $T_{ai}\gg{}T_e$
\cite{Mor15} (as in the calculations of Ref. 2) assumes a moderate
degree of ionization of the unperturbed chromospheric gas (ahead of
the front); here $T_{ai}$ is the atom-ion temperature of the
plasma.{\footnote{We note that: (a) the authors of Ref. 2, as
opposed to Ref. 10, have incorrectly calculated the probability of
escape $\theta_{12}$ of a resonance photon beyond the confines of
the plasma without scattering (substituting the Doppler optical
depth in an expression obtained for a Holtsmark profile); (b) in
Ref. 10, in the case of emergence of a photon from the center of a
plane layer
$\theta_{12}^c\approx0.6{}\overline{\theta}_{12}<\overline{\theta}_{12}$,
where
$\overline{\theta}_{12}$ is the value of $\theta_{12}$ averaged over the layer (Eq. (55) in Ref. 10).}}\\
{\small{\bfseries{3. Neutral helium lines.}} For high velocities
$v_{sh}$ of the {\it non-stationary} chromospheric shock wave
propagating in a partially ionized plasma in the direction of the
photosphere, this increase in the electron temperature makes it
possible to create conditions for ionization of helium atoms
($\mathrm{HeI}$) and excitation of their discrete levels by electron
impact (radiative cooling is substantially non-stationary
\cite{Bychkov19}, so the degree of ionization and the excitation
state of the atoms are determined not only by the current value of
$T_e$); this is a known effect \cite{Kor72} in the theory of {\it
stationary} shock waves with emission. In this case the subsequent
de-excitation by allowed spontaneous transitions can ensure
\cite{Mor19} the appearance and enhancement of {$\mathrm{HeI}$}
lines in the spectra of powerful solar and stellar flares ({\it
e.g.}, lines with $\lambda=10830\text{\,\AA}$ \cite{Schmidt}). We
recall, however (see Ref. 15), that $v_{sh}$ is increased by raising
the energy flux $F_0$ in the beam of accelerated electrons.

We emphasize that, because of the rapid radiative cooling of the
dense gas behind the shock front ({\it{e.g.}}, Ref. 1) the situation
described above: (a) corresponds to zones in which the plasma
relaxes to a state of thermal equilibrium for a certain {\it range}
of velocities $v_{sh}$ (this effect can be illustrated in a model of
a ``set'' of stationary shocks with emission \cite{Bychkov19}); (b)
is detected for heating pulses that partially overlap in time and
act on different segments of the upper chromosphere of the Sun and
dMe stars.\\
{\bf 4. $\mathrm{H}_{\alpha}$ lines with ``blue'' asymmetry in the
wings.} It is known that in the spectra of some flares on dMe stars
long-lived asymmetric profiles of the $\mathrm{H}_\alpha$ lines are
observed with enhanced intensity in the blue wing rather then the
red ({\it{e.g.}}, Refs. 16 and 6). Honda {\it et al.} \cite{Honda}
noted a possible relationship between the formation of radiation of
this sort and the presence in the flare atmosphere of a red dwarf of
cold ($T\sim10^4{\text{ K}}$ \cite{Honda}) gas. At the same time,
the existence  of broad wings in $\mathrm{H}_{\alpha}$ (Fig. 7, Ref.
16) is evidence (Eason {\it et al.} \cite{Eason}, Fig. 9; an
approximation of part of the red wing of the line by a Stark
profile) of a {\it fairly high} gas density that is responsible for
their formation. (We note that the value of $\log{n_e}$ is found in
Ref. 16 without accounting for the significant optical depth of
$\mathrm{H}_\alpha$ in the core (the characteristic ``dip'' in the
center) that is indicative of possible opacity of this line even in
the wings.) Using $\Delta\lambda_D$ (the Doppler half width) and
$\log{n_e}$ found in Ref. 16, Morchenko {\it et al.}
\cite{Mor19,Mor15} showed (section 5.3 in \cite{Mor15}; introduction
to \cite{Mor19}) that the $\mathrm{H}_\alpha$ profile in the
observations \cite{Eason} is approximately described by a model with
a ``Doppler core'' and ``Stark-broadened wings.''

Honda {\it et al.} \cite{Honda} also give possible explanations of
the nature of the ``blue'' asymmetry in the wings of
$\mathrm{H}_\alpha$ known from the physics of solar flares. These
explanations, however, are pointed out as insufficient by the
authors \cite{Honda}.

Given these remarks, there is some interest in interpreting the
nature of the blue asymmetry in the wings of $\mathrm{H}_\alpha$
based on the existence of a {\it velocity field} of gas in a
chromospheric condensation and in thermal relaxation zones
\cite{Kost76}. According to this point of view, ``the type of
asymmetry is determined by the {\it sign} of the velocity gradient
in that region where the intensity of the central part of the line
is formed'' (a red asymmetry corresponds to expansion of this region
and a blue, to its compression \cite{Kost76}); here the resulting
line profile shows up \cite{Kost76} as a superposition of profiles
that take into account the non-uniformity of the physical parameters
of the beams of accelerated electrons (see Fig. 3 of Ref. 17). This
interpretation is favored by the absence of a significant Doppler
shift in the center of the $\mathrm{H}_\alpha$ line in observational
data \cite{Eason,Honda}.\\
{\bf 5. Emission in the ``white'' light continuum.} The author,
therefore, proposes the following scheme for the locations of the
plasma ``layers'' responsible for the continuum (in white light) and
line radiation during the impulsive phase of powerful flares on dMe
stars:

(a) a gas that is {\it transparent} in the Balmer continuum is
formed mainly in a dynamically inhomogeneous chromospheric
condensation (the optical depth at a wavelength of $4170\text{
\AA}$, $\tau_{4170}\ll1$, as in Refs. 18 and 19) and also in thermal
relaxations zones\footnote{With a geometric thickness that is
considerably less than the thickness of the chromospheric
condensation toward the end of the heating.} and in the region {\it
ahead} of a non-stationary chromospheric shock wave (the precursor
and heating by high-energy, $E\gg{}E_{10}$ ($E_{10}$ is the cutoff
on the low energy side), electrons from a beam with a decaying
power-law spectrum). This point of view agrees qualitatively (for
the chromospheric condensations and the thermal relaxations zones
--- there is a substantial Balmer jump) with the results of an
analysis of the spectra of the flare YZ CMi (dM4.5e) in the NUV
wavelength range (from 3350 to $\lambda<3646\text{ \AA}$) by
Kowalski {\it et al.} \cite{Kow10} (Fig. 1c in \cite{Kow10}, the
response of the chromosphere of the red dwarf to heating by a beam
of non-thermal electrons with an energy flux of $F_0=10^{11}\text{
erg}/{\text{cm}^2\text{s}}$);

(b) the quasi-black-body ``blue'' (at the maximum brightness of
powerful flares, $T_{bb}\sim10^4\text{ K}$), $\lambda$ from 4000 to
4800 \AA, and red (in the slow decay phase) components of the
optical continuum arise near the photosphere (Grinin and Sobolev
\cite{Grin77});

(c) the HI emission lines are mainly localized in the chromospheric
condensation and zones of thermal relaxation. Here the populations
of the atomic levels and the degree of ionization of the plasma are:
(a) close to the equilibrium values in a chromospheric condensation
with sufficient geometric thickness\footnote{The non-stationary
shock front has substantially ``overtaken'' the temperature jump.}
(Kowalski and Allred \cite{Kow18}, Morchenko {\it et al.}
\cite{Mor15,Mor19}) and in the region where the blue continuum is
formed (at the brightness maximum of powerful flares)
\cite{Grin77,Mor16}; (b) differ sharply \cite{Kow18} from the
equilibrium values in the thermal relaxation zones behind the front
of the non-stationary chromospheric shock wave (according to the
general rule the deviations for the populations of the atomic levels
$\nu_k$ decrease with increasing principal quantum number $k$). The
gas radiating behind the front of the non-stationary chromospheric
shock wave is stable with respect to radiative cooling (this
conclusion follows from the calculations of Ref. 2). The dense
plasma of the chromospheric condensation, which is near a state of
complete thermalization, ensures (see the review \cite{Grin84}) the
{\it gently sloping} and inverse Balmer decrements that are
characteristic ({\it e.g.}, \cite{Gersh77}) of the brightness maxima
of flares on red dwarfs.

The following observational data confirm the validity of this
scheme:

(a) the time ``evolution'' of the Balmer continuum closely follows
the corresponding time variations in the atomic hydrogen Balmer
lines (Kowalski {\it et al.} \cite{Kow10}, bottom part of Fig. 1d).
This fact indicates some relationship between their formation;

(b) ``the Balmer continuum manifests a slow decay ..., which is
typical of emission in the Balmer series'' lines of the hydrogen
atom in the spectra of flares \cite{Gersh77,Kow10};

(c) during the time a flare is decaying, the emission region in the
Balmer continuum is ``always considerably ($\sim3-16$ times)
greater'' \cite{Kow10} than the region occupied by the component
with $T_{bb}\sim10^4\text{ K}$. This result can be interpreted as,
on one hand, an increase in the geometric thickness of the region
between the temperature jump and the front of the non-stationary
shock wave with emission (during the time of impulsive heating the
thermal relaxation zone ``travels'' into a region of ever higher
values of the Lagrangiane coordinate $\xi$ (see Fig. 3 in Ref. 18),
while the temperature jump is essentially motionless relative to the
shock front) and, on the other hand, as a transition of part of the
plasma in the near-photospheric layers into a state in which the
spectrum of the radiation is {\it transparent} beyond the Balmer
jump;

(d) the time evolution of the radiation in the $U$ band differs
(\cite{Kow10}, Fig. 1d) from the behavior of the Balmer continuum,
with the largest deviations corresponding \cite{Kow10} to the {\it
peaks} in the light curve ({\it{e.g.}}, at $t\sim130$ min), when the
flare spectrum is dominated \cite{Zhil,Kow10} by the
quasi-black-body ``blue'' component of the optical continuum. Thus,
these components of the radiation in white light are formed in
different regions of the chromosphere;

(e) during the slow decay phase of flares, in the region beyond the
Balmer jump a larger range of wavelengths is ``occupied'' by the
``red'' component of the optical continuum (Fig. 31 in
\cite{Kow13}). This result is explained by the fact that as the
hydrogen plasma undergoes a transition into a state close to LTE
(the quasi-black-body spectrum), the {\it red} part of the spectrum
thermalizes initially (Morchenko {\it et al.} \cite{Mor15}, Fig. 2),
and then the blue.

Furthermore, the {\it increase} in the energy flux $F_0$ for fixed
$E_{10}$ and spectral index (indices) of the non-thermal electrons
leads to a {\it reduction} in the ``evolution'' time of the
chromospheric condensation and the thermal relaxation zones. From a
physical standpoint this happens because (a) the higher values of
$F_0$ correspond to higher velocities $v_{sh}$, so that the shock
front ``moves'' into the depth of the chromosphere of dMe stars more
rapidly; (b) large $v_{sh}$ causes stronger compression of the gas
behind the non-stationary shock front (for an ideal monatomic gas
with a constant specific heat the limiting compression is 4), and,
thereby, more rapid ({\it e.g.}, Ref. 1) radiative cooling. Hence,
the use of extremely high, $F_0=10^{13}\text{
erg}/{\text{cm}^2\text{s}}$ (Kowalski {\it et al.} \cite{KowIAU}),
energy fluxes in gas dynamic calculations of the type in Ref. 18
lead to a {\it shortening} of the ``lifetime'' of the Balmer lines,
which is inconsistent with the need to interpret their {\it
long-term} emission \cite{Gersh77,Kow10} in the spectra of flares.

We note that in the proposed scheme, the gas generating the
quasi-black-body radiation (at the maximum brightness of powerful
flares) lies much deeper than the plasma responsible for the line
spectrum and the spectrum transparent in the Balmer continuum (a
positive gradient of the density in the perturbed chromosphere of a
red dwarf). In addition, this sort of scheme starts with moderately
high, from $\sim10^{11}\text{ erg}/{\text{cm}^2\text{s}}$ to
$\sim10^{12}\text{ erg}/{\text{cm}^2\text{s}}$ $F_0$ (the return
current problem), and is consistent (Tsap {\it et al.} \cite{Tsap})
with the standard model of solar flares, since it does not assume
the existence (see the references in \cite{Tsap}) of dense
($n_e>10^{12}\text{ cm}^{-3}$) flare coronal loops.\footnote{In the
calculations \cite{Katsova} the contribution of soft x-rays to
heating the dense layers of the chromosphere ($\xi\sim10^{21}$
cm${}^{-2}$ \cite{Katsova}) is accounted for in the heating function
$P_e(\xi)$ by taking the coefficient $\beta=1$.}\\
{\bf 6. Conclusion.} In the author's opinion, a theoretical check of
the agreement between this scheme and observations of flares will
involve: (a) a study of mechanisms for accelerating particles to
high energies in the model \cite{Grin77} and (b) a study of the
effect of the radiation field of the chromospheric plasma (the gas
with $T\sim10^7\text{ K}$ \cite{Katsova}, the thermal wave; the
thermal relaxation zones; the gas ahead of the non-stationary shock
wave) on the deep (near-photospheric) layers of flare (dMe) stars.}

{\small

\newpage
}
\end{document}